\documentclass[conference]{IEEEtran}
\usepackage{graphicx}
\usepackage{caption}
\usepackage{subcaption}

\usepackage{longtable}
\usepackage{stfloats}
\usepackage{graphicx}

\usepackage{booktabs}
\usepackage{multirow}
\usepackage{tabularx}

\usepackage{ifpdf}
\usepackage[T1]{fontenc}
\usepackage{amssymb}
\usepackage{cite}

\ifCLASSINFOpdf
   \usepackage{graphicx}
	\usepackage{epstopdf}
   \graphicspath{{images/}}   
   \graphicspath{{../pdf/}{../jpeg/}{../eps/}{../png/}}
   \DeclareGraphicsExtensions{.pdf,.jpeg,.png,.eps}
\else
   \graphicspath{{../images/eps/}}
   \DeclareGraphicsExtensions{.eps}
\fi
\graphicspath{{images/}}

\usepackage[cmex10]{amsmath}
\usepackage{multirow}

\usepackage{multirow}

\usepackage{booktabs}
\usepackage{multicol}

\usepackage[shortcuts,acronym, nomain]{glossaries}

\makeglossaries 

\newacronym{v2x}{V2X}{Vehicle-to-everything}
\newacronym{v2v}{V2V}{Vehicle-to-Vehicle}
\newacronym{v2i}{V2I}{Vehicle-to-Infrastructure}
\newacronym{v2p}{V2P}{Vehicle-to-Pedestrian}
\newacronym{eebl}{EEBL}{Emergency Electronic Brake Assist}
\newacronym{bsw}{BSW}{Blind Spot Warning}
\newacronym{dsa}{DSA}{Dynamic Speed Advisory}
\newacronym{dsrc}{DSRC}{Dedicated Short Range Communications}
\newacronym{cv2x}{C-V2X}{Cellular-V2X}
\newacronym{fcc}{FCC}{Federal Communications Commission}
\newacronym{bler}{BLER}{Block Error Rate}
\newacronym{per}{PER}{Packet Error Rate}
\newacronym{snr}{SNR}{Signal-to-Noise Ratio}
\newacronym{astm}{ASTM}{American Society for Testing and Materials}
\newacronym{ofdm}{OFDM}{Orthogonal Frequency Division Multiplexing}
\newacronym{wave}{WAVE}{Wireless Access in Vehicular Environments}
\newacronym{ppdu}{PPDU}{Protocol Packet Data Unit}
\newacronym{ltf}{LTF}{Long Training Field}
\newacronym{stf}{STF}{Short Training Field}
\newacronym{sig}{SIG}{Signal}
\newacronym{psdu}{PSDU}{PLCP Service Data Unit}
\newacronym{mcs}{MCS}{Modulation-Coding Scheme}
\newacronym{3gpp}{3GPP}{3rd Generation Partnership Project}
\newacronym{ieee}{IEEE}{Institute of Electrical and Electronics Engineers}
\newacronym{mac}{MAC}{Medium Access Control}
\newacronym{wlan}{WLAN}{Wireless Local Area Network}
\newacronym{cits}{C-ITS}{Cooperative Intelligent Transportation Systems}
\newacronym{d2d}{D2D}{Device-to-Device}
\newacronym{ue}{UE}{User Equipment}
\newacronym{rp}{RP}{Resource Pool}
\newacronym{sl}{SL}{SideLink}
\newacronym{scfdma}{SC-FDMA}{Single Carrier Frequency Division Multiple Access}
\newacronym{papr}{PAPR}{Peak to Average Power Ratio}
\newacronym{stch}{STCH}{Sidelink Traffic Channel}
\newacronym{sbbch}{SBBCH}{Sidelink Broadcast Control Channel}
\newacronym{slsch}{SL-SCH}{Sidelink Shared Channel}
\newacronym{pssch}{PSSCH}{Physical Sidelink Shared Channel}
\newacronym{pscch}{PSCCH}{Physical Sidelink Control Channel}
\newacronym{sci}{SCI}{Sidelink Control Message}
\newacronym{prb}{PRB}{Physical Resource Block}
\newacronym{ul}{UL}{Uplink}
\newacronym{dl}{DL}{Downlink}
\newacronym{enb}{eNB}{eNodeB}
\newacronym{dmrs}{DMRS}{Demodulation Reference Symbol}
\newacronym{c-its}{C-ITS}{Cooperative Intelligent Transportation Systems}
\newacronym{ima}{IMA}{Intersection Movement Assist}
\newacronym{dnpw}{DNPW}{Do Not Pass Warning}
\newacronym{rsu}{RSU}{Road Side Unit}
\newacronym{dft}{DFT}{Discrete Fourier Transformation}
\newacronym{cp}{CP}{Cyclic Prefix}
\newacronym{los}{LOS}{Line of Sight}
\newacronym{nlos}{NLOS}{Non-Line of Sight}
\newacronym{itu}{ITU}{International Telecommunication Union}
\newacronym{eva}{EVA}{Extended Vehicular A}
\newacronym{va}{VA}{Vehicular A}
\newacronym{vb}{VB}{Vehicular B}
\newacronym{phy}{PHY}{Physical Layer}
\newacronym{awgn}{AWGN}{Additive White Gaussian Noise}
\newacronym{nxid}{NXID}{V2X Scrambling Identity}
\newacronym{sc}{SC}{Sub-Carrier}
\newacronym{sbcch}{SBCCH}{Sidelink Broadcast Control Channel}
\newacronym{slbch}{SLBCH}{Sidelink Broadcast Channel}
\newacronym{psbch}{PSBCH}{Physical SideLink Broadcast Channel}

\begin{document}

\title{Link Level Performance Comparison of C-V2X and ITS-G5 for Vehicular Channel Models}
%
\author{\IEEEauthorblockN{Raja Sattiraju, Donglin Wang, Andreas Weinand and Hans D. Schotten}
\IEEEauthorblockA{Chair for Wireless Communication \& Navigation \\
University of Kaiserslautern\\
\{sattiraju, dwang, weinand, schotten\}@eit.uni-kl.de}}
\maketitle

\begin{abstract}
	\ac{v2x} communications plays a significant role in increasing traffic safety and efficiency by enabling vehicles to exchange their status information with other vehicles and traffic entities in their proximity. In this regard, two  technologies emerged as the main contenders for enabling \ac{v2x} communications which have stringent requirements in terms of latency and reliability due to their apparent safety criticality. The first one is the \ac{dsrc} standard (referred to as ITS-G5 in Europe) that is well researched since 20 years and has attained enough technical maturity for current deployment. The second one is the relatively new \ac{cv2x} standard that is nevertheless, based on the \ac{3gpp} standard family that have  successful deployments in almost every corner of the globe. In this work, we compare the link level performance of the \ac{phy} protocols for both the technologies for different vehicular fading channel models. To this end, we construct and simulate the \ac{phy} pipelines and show the performance results by means of \ac{bler} versus \ac{snr} graphs. Our investigations show that \ac{cv2x} performs better than ITS-G5 for almost all the considered channel models due to better channel coding and estimation schemes.

\end{abstract}

\section{Introduction}
5G networks enable native support for new vertical domains such as vehicular and industrial communications. \ac{v2x} communication encompasses any form of communication between a vehicle and surrounding traffic entities and it includes different modes such as \ac{v2v}, \ac{v2i} and \ac{v2p}. \ac{v2x} has the potential to significantly decrease traffic accidents and at the same time, increases traffic efficiency. Examples of safety applications include \ac{eebl}, \ac{bsw} etc. while Platooning and \ac{dsa} etc. are some example applications for increasing traffic efficiency.

\ac{v2x} applications also bring with them very stringent requirements in terms of latency and reliability due to their apparent safety criticality. Added to this is the profound unpredictability of wireless channels at highly dynamic mobile scenarios such as driving on highways. If these challenges are not addressed properly, the benefits of \ac{v2x} cannot be exploited and utilized. Therefore, a lot of research has been done in order to design robust \ac{phy} layer protocols that can effectively combat the channel variations in vehicular communication scenarios. This resulted in two standards namely \ac{dsrc} (ITS-G5 in Europe) that is based on \ac{ieee} 802.11 \ac{wlan} standard and \ac{cv2x} that is based on \ac{3gpp} standards. Though these \ac{v2x} \ac{phy} standards are accompanied with their own enhancements at higher layers, we limit our discussion in this paper to  the \ac{phy} layer.

The development of any new wireless standard necessitates the use of simulation in order to evaluate and test the proposed standard. In this paper, we evaluate the \ac{phy} layer of both ITS-G5 and \ac{cv2x} in terms of link level performance under various \ac{v2v} channel models. Some of these channel models were proposed by \ac{itu} and others were derived by means of  measurement campaigns. Using extensive link-level simulations, we compare the \ac{snr} versus \ac{bler} performance for both the technologies.

\subsection{Related State of the Art}
The underlying \ac{phy} for ITS-G5 is based on \ac{ieee} 802.11p standard, a well matured technology that has been researched for over 20 years. Hence, its \ac{phy} layer performance has been evaluated in many works notably \cite{Tan2008, Fernandez2012, Paier2010}. In \cite{Mir2014} and \cite{Moller2014}, the performance of \ac{ieee} 802.11p has been compared with legacy LTE networks (no sidelink) for different \ac{los}/\ac{nlos} scenarios. The link level performance of Release.12 LTE \ac{d2d} is done in \cite{Cintron2018}. \ac{cv2x} is relatively new with the first specification released in 2016. Since then, there has been some works that compared the performance of both the technologies. In \cite{Hu2017}, the authors compared both the technologies in terms of \ac{per} using  WINNER II channel model \cite{Kyosti}. However, these models are suitable for only base station to mobile \ac{ue} links and do not explicitly consider \ac{v2v} channel models. In \cite{Mannoni2019}, the authors compared the performance of both the technologies for \ac{itu}-\ac{eva} channels. However, it provides no results for other \ac{itu} \ac{v2v} channel models. In contrast, our work considers a broad spectrum of \ac{v2v} channels from \ac{itu} (\ac{va}, \ac{vb} and \ac{eva}) \cite{IEEE} and also the models derived from field measurements in \cite{Alexander2011a}. 

The rest of the paper is organized as follows. Section II outlines the fundamentals of both ITS-G5 and \ac{cv2x}. In section III, we present the baseband processing pipeline for both the technologies. The \ac{v2v} fading channel models are presented in Section IV. It also presents the \ac{snr} versus \ac{bler} graphs for the considered channel models along with some discussions. Section V concludes the paper with a summary of the results.

\section{Candidate Technologies}
This section provides a brief overview of the \ac{phy} layer for both ITS-G5 (also referred to as \ac{dsrc} in US) and \ac{cv2x}.

\subsection{DSRC}
The genesis of \ac{dsrc} can be traced back to 1999 when the US \ac{fcc} granted 75 MHz of dedicated bandwidth in 5.9 GHz region for automotive applications. In 2002, on the basis of extensive research and testing, the \ac{astm} published \ac{astm} E2213 standard that recommended that the candidate be based on a modified version of \ac{ieee} 802.11a \cite{IEEEComputerSociety.LAN/MANStandardsCommittee.2012}. This led to the formation of an \ac{ieee} study group that drafted an amendment based on \ac{astm} recommendation and named it \ac{ieee} 802.11p. Similar to \ac{ieee} 802.11a, \ac{ieee} 802.11p uses \ac{ofdm} at \ac{phy} along with re-using the same preamble and pilot design for synchronization and channel estimation. The only difference is that \ac{ieee} 802.11p operates in \textit{half-clocked} mode halving the 20 MHz channel spacing to 10 MHz and effectively doubling the symbol timing. This enables \ac{ieee} 802.11p to better handle the high mobility scenarios as compared to \ac{ieee} 802.11a. Furthermore, over the top protocols by \ac{wlan} and 1609 \ac{dsrc} working group complemented \ac{ieee} 802.11p to enable \ac{wave} and these whole set of standards are referred to as \ac{dsrc}.

The \ac{ieee} 802.11p equivalent in the European \ac{c-its} stack covering \ac{phy} and MAC layers is termed as ITS-G5 \cite{Festag2014}. Similar to \ac{dsrc}, it also operates in the 5.9 GHz band using \ac{ofdm} at the same half-clocked mode but with the adapted spectrum masks. Even though the underlying network protocol is based on IPv6, \ac{c-its} specifies an additional multi-hop routing protocol called Geo-networking that uses geographical coordinates for addressing and forwarding messages. Geo-networking is optimized for multi-hop communications with geo-addressing, providing enhanced support for applications albeit at an increased protocol complexity and overhead.

\begin{figure}[h]
	\centering
	\includegraphics[width=0.49\textwidth]{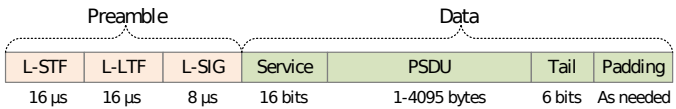}
	\caption{Packet Structure - \ac{ieee} 802.11p}
	\label{packet_11p}
\end{figure} 

The data structure that the \ac{phy} receives from MAC in \ac{ieee} 802.11p is termed as \ac{ppdu} (or PSDU) that is made up of three components - preamble and data fields as shown in Figure \ref{packet_11p}. In the preamble, \ac{stf} is used for packet detection, coarse frequency correction and automatic gain control. \ac{ltf} is used for fine frequency correction, fine symbol timing offset correction and pilot based channel estimation. The \ac{sig} field contains packet information for the received configuration such as \ac{mcs} used and the \ac{psdu} length. The service field consists of 16 zeros to initialize the data scrambler. \ac{psdu} contains the actual user data. Tail bits are used to terminate the convolutional code and the padding bits are added to ensure an integer number of symbols.


For actual transmission, \ac{ofdm} is used with a total of 64 \acp{sc}. Out of these 64 \acp{sc}, 52 are used for carrying data and pilot symbols and the remaining 12 are null \acp{sc} that carry no data. The null \acp{sc} occupy the central 11 \acp{sc} and the 0th \ac{sc}. The pilot symbols occupy 4 \acp{sc} with indices 7, 10, 44 and 58. The remaining 48 \acp{sc} are used for data \cite{Abdelgader2014}. The actual length of data depends on the choice of \ac{mcs} with the supported schemes outlined in Table \ref{tab:mcs_11p}

\begin{table}[]
	\centering
	\caption{MCS Schemes - \ac{ieee} 802.11p}
	\label{tab:mcs_11p}
	\resizebox{0.49\textwidth}{!}{%
		\begin{tabular}{@{}ccccc@{}}
			\toprule
			\multicolumn{1}{l}{\textbf{MCS}} & \multicolumn{1}{l}{\textbf{Modulation}} & \multicolumn{1}{l}{\textbf{Coding Rate}} & \multicolumn{1}{l}{\textbf{\begin{tabular}[c]{@{}l@{}}Coded bits per \\ OFDM Symbol\end{tabular}}} & \multicolumn{1}{l}{\textbf{\begin{tabular}[c]{@{}l@{}}Data Rate\\ (MBPS)\end{tabular}}} \\ \midrule
			0 & BPSK & 1/2 & 48 (24 data bits) & 3 \\
			1 & BPSK & 3/4 & 48 (36 data bits) & 4.5 \\
			2 & QPSK & 1/2 & 96 (48 data bits) & 6 \\
			3 & QPSK & 3/4 & 96 (72 data bits) & 9 \\
			4 & 16QAM & 1/2 & 192 (96 data bits) & 12 \\
			5 & 16QAM & 3/4 & 192 (144 data bits) & 18 \\
			6 & 64QAM & 2/3 & 288 (192 data bits) & 24 \\
			7 & 64QAM & 3/4 & 299 (216 data bits) & 27 \\ \bottomrule
		\end{tabular}%
	}
\end{table}

\subsection{\acf{cv2x}}
\ac{3gpp}'s Release.12 standard included significant changes to the legacy LTE architecture by introducing the concept of direct \ac{d2d} communications. Known collectively as Proximity Services (ProSe), this mode enables \acp{ue} that are in close proximity to directly establish a communication link (via a PC5 interface) between themselves instead of relying on the network infrastructure. Cellular resources in the \ac{ul} are used for ProSe services mainly because of two reasons: 1) \ac{ul} transmissions are sporadic compared to \ac{dl} where the eNB has always something to transmit and 2) Due to the low transmission power and geographical separation of the \acp{ue}, interference is also less in the \ac{ul} band.

%

\subsubsection*{V2X Enhancements}
The LTE \ac{d2d} standard is proposed keeping in mind the emergency public communications and proximity based advertisements using conventional \acp{ue}, i.e., smartphones, whose positions are usually assumed to be semi-static. However, \ac{v2x} links are highly dynamic with higher channel uncertainties. Secondly, the node density is also comparatively higher especially in urban areas. Hence, to this end \ac{3gpp} introduced few fundamental modifications to the PC5 interface (sidelink interface) to meet the more stringent latency and reliability requirements associated with the vehicular use cases. They are

\begin{enumerate}
	\item[i.] Using additional \acp{dmrs} (4 instead of 3) to handle the higher Doppler corresponding to relative speeds of up to 500 km/h and at high frequency (5.9 GHz ITS band)
	
	\item[ii.] Using a new resource scheduling assignment of \ac{ul} resources where the control data and the shared data are transmitted in a single subframe over adjacent \acp{prb}. More information about the concept of \acp{rp} can be found in the next section.
	
	\item[iii.] For out of coverage resource scheduling assignment, a sensing with semi-persistent transmission based mechanism was introduced. Since \ac{v2v} traffic is mostly periodic in nature, this property is utilized to sense congestion on a resource and estimate future congestion on that resource.
\end{enumerate}

\subsubsection*{\acf{rp}}
In contrast to \ac{ieee} 802.11p that use the entire available bandwidth (10 MHz) for each packet transmission, the sidelink transmissions are scheduled to operate side by side with the \ac{ul} transmissions and only in a subset of \acp{sc}. Hence, new measures for resource allocation and transmission scheduling are required. This is achieved by means of \acp{rp}; a set of resources assigned to the \ac{sl} operation. It consists of a set of sub-frames and resource blocks within. The physical resources (sub-frames and resource blocks) associated with a given pool are partitioned into a sequence of repeating \textit{hyperframes} known as \ac{pssch} periods, also referred to as the Scheduling Assignment (SA) period or Sidelink Control (SC) period. Within a \ac{pssch} period there are separate sub-frame pools and resource block pools for control and data. The \ac{pscch} carries \ac{sci} messages, which describe the dynamic transmission properties of the \ac{pssch} that follow it. The receiving \ac{ue} searches all configured \ac{pssch} resource pools for \ac{sci} transmissions of interest to it.

\begin{figure}[h]
	\centering
	\includegraphics[width=0.49\textwidth]{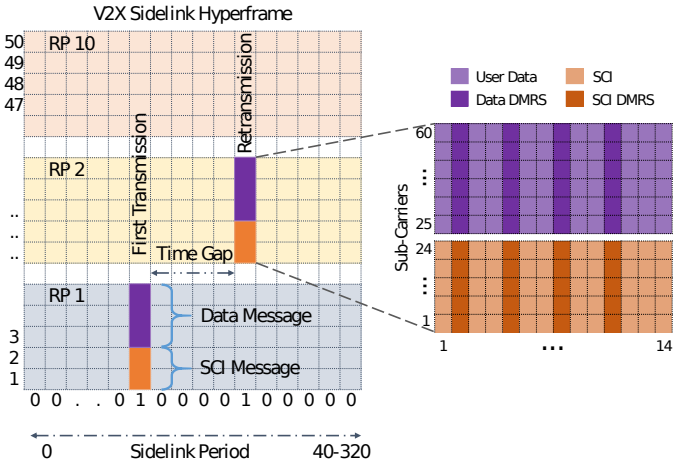}
	\caption{Example V2X Hyperframe}
	\label{v2x_hyperframe}
\end{figure} 

Figure \ref{v2x_hyperframe} illustrates an example sidelink hyperframe for a bandwidth of 10 MHz and a \ac{pssch} period of 40 ms. Within a \ac{pssch} period, the actual sidelink transmissions can be found on any two subframes (for first transmission and retransmission) given by the subframe bitmap. For the considered bandwidth of 10 MHz, there are 50 \acp{prb} that are divided into 10 sub-pools each consisting of 5 contiguous \acp{prb}. A \ac{ue} can use one or multiple sub-pools for transmission as specified by higher layer messages. For retransmission (1 blind retransmission is supported by default), the \ac{ue} can use the same set of sub-pools as the first transmission and use different sub-pools for the subsequent retransmission. In our example, the \ac{ue} uses \ac{rp}1 for the first transmission and \ac{rp}2 for the retransmission.

The \ac{sci} message always spans 2 \acp{prb} which is succeeded by the data message. For the given example, a data message spanning over 3 \acp{prb} is assumed. The content of each message is also illustrated in Figure \ref{v2x_hyperframe}. In line with the LTE specification, each \ac{prb} consists of 12 \acp{sc} in the frequency domain and 14 \ac{ofdm} symbols in the time domain. Symbols 2, 5, 8 and 11 are used for transmitting \ac{dmrs} that are used for frequency correction and channel estimation. The remaining 10 symbols are used to carry the actual data.


\begin{table}[]
	\centering
	\caption{C-V2X MCS Schemes}
	\label{tab:cv2x_mcs}
	\resizebox{0.35\textwidth}{!}{%
		\begin{tabular}{@{}cccc@{}}
			\toprule
			\textbf{MCS Index} & \textbf{Modulation} & \textbf{\begin{tabular}[c]{@{}c@{}}Transport \\ Block Size\end{tabular}} & \textbf{\begin{tabular}[c]{@{}c@{}}Effective \\ Coding Rate\end{tabular}} \\ \midrule
			0 & QPSK & 1320 & 0.127 \\
			1 & QPSK & 1736 & 0.167 \\
			2 & QPSK & 2152 & 0.207 \\
			3 & QPSK & 2792 & 0.269 \\
			4 & QPSK & 3496 & 0.337 \\
			5 & QPSK & 4264 & 0.411 \\
			6 & QPSK & 4968 & 0.479 \\
			7 & QPSK & 5992 & 0.577 \\
			8 & QPSK & 6712 & 0.647 \\
			9 & QPSK & 7480 & 0.721 \\
			10 & QPSK & 8504 & 0.820 \\
			11 & 16QAM & 8504 & 0.410 \\
			12 & 16QAM & 9528 & 0.459 \\
			13 & 16QAM & 11064 & 0.533 \\
			14 & 16QAM & 12216 & 0.589 \\
			15 & 16QAM & 13536 & 0.652 \\
			16 & 16QAM & 14688 & 0.708 \\
			17 & 16QAM & 15840 & 0.763 \\
			18 & 16QAM & 17568 & 0.857 \\
			19 & 16QAM & 19080 & 0.920 \\
			20 & 16QAM & 20616 & 0.994 \\ \bottomrule
		\end{tabular}%
	}
\end{table}

The \ac{phy} layer of the \ac{cv2x} is same as the LTE uplink and uses \ac{scfdma} as the access technique. \ac{scfdma} has lower \ac{papr} when compared to \ac{ofdm} while at the same time combining the advantages of multipath interference resilience and flexible sub-carrier frequency allocation that \ac{ofdm} provides. The individual \acp{sc} are modulating using one of the three modulation schemes namely - QPSK, 16-QAM and 64-QAM. Table \ref{tab:cv2x_mcs} outlines the different \ac{mcs} schemes for a bandwidth of 10 MHz \cite{TSGR2018b}.\footnote{Before \ac{scfdma} modulation, the last symbol is set to 0 in accordance with 3GPP specification. Therefore the total useful symbols per subframe becomes 9. These values are used for calculating the effective coding rate.}

\section{Link level Simulation Methodology}
For link level simulation, the complete transmit and receive operations needs to be built. This section outlines the baseband processing for both \ac{ieee} 802.11p and \ac{cv2x}

\subsection{\ac{ieee} 802.11p Baseband Processing}

\begin{figure}[h]
	\centering
	\includegraphics[width=0.49\textwidth]{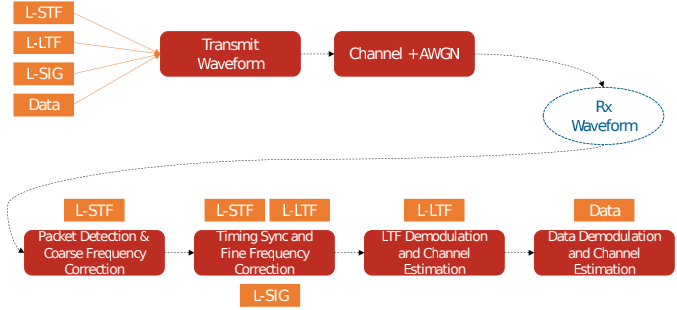}
	\caption{\ac{ieee} 802.11p Simulation Pipeline}
	\label{11p_pipeline}
\end{figure}

The simulation pipeline for \ac{ieee} 802.11p is outlined in \ref{11p_pipeline}. The \ac{ltf}, \ac{stf} and \ac{sig} symbols are concatenated together to form the preamble. The user data is convolutionally encoded and mapped to symbols corresponding to the selected \ac{mcs}. Finally, the preamble and the data symbols are concatenated together and \ac{ofdm} modulated to create the time-domain waveform. The waveform is passed through a fading channel and \ac{awgn} noise is added to it to get the received waveform. The following operations are performed sequentially on the received waveform to decode the data


\begin{enumerate}
	\item[i.] Packet detection, estimation of coarse packet offset and coarse frequency correction using the \ac{stf}
	\item[ii.] Fine packet offset estimation, fine frequency offset correction and fine symbol timing offset correction using the complete preamble
	\item[iii.] Demodulation of \ac{ltf} and channel estimation using the pilot symbols
	\item[iv.] The constructed channel coefficient matrix is used to demodulate, equalize and decode the user data
\end{enumerate} 

\subsection{\ac{cv2x} Baseband Processing}
\begin{figure}[h]
	\centering
	\includegraphics[width=0.49\textwidth]{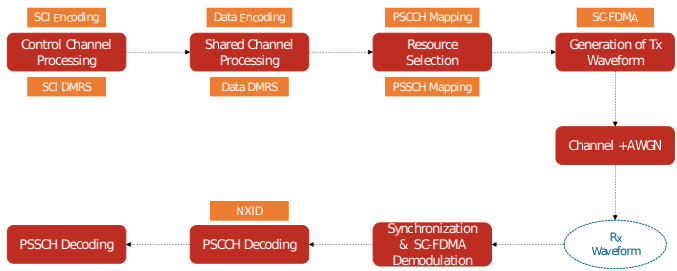}
	\caption{C-V2X Simulation Pipeline}
	\label{cv2x_pipeline}
\end{figure}

Figure \ref{cv2x_pipeline} shows the baseband processing pipeline for \ac{cv2x}. As it can be seen, control message and data message processing is done separately and these symbols are sequentially added to the time-frequency resource grid.

\subsubsection{Control Channel Processing}
The first step is control channel processing is to create and encode the \ac{sci} message. For \ac{v2x} transmission a 'Format 1' \ac{sci} message is generated that consists of information such as the \ac{mcs} , Resource Indication Value (RIV), the time gap between initial transmission and retransmission and the retransmission index (0 in case of initial transmission and 2 in case of first retransmission).  The generated binary message is encoded using a convolutional encoder followed by rate matching, interleaving and a 16-bit CRC is attached to the encoded message. The 16-bit CRC is then converted into a decimal and this value is referred to as \ac{nxid}. It is used as the initialization value for generating the gold sequence which is in turn used for scrambling the user data. This effectively means that the receiver would be able to decode the data message if and only if it has decoded the \ac{sci} message successfully and recovered the 16 bit CRC remainder.

After generating the binary code word, next processing steps involve \ac{pscch}-specific scrambling, QPSK modulation and \ac{scfdma} transform precoding to generate symbols. The generated PSCCH symbols (240) and are cyclic shifted with a random value chosen from set [0, 3, 6, 9] in order to reduce the effect of interference. Finally, 4 \ac{dmrs} symbols are generated and mapped to the remaining 4 time domain symbols ([2,5,8,11]).

\subsubsection{Shared Channel Processing} 
\ac{slsch} processing includes type-24A CRC calculation, code block segmentation (including type-24B CRC attachment, if present), turbo encoding, rate matching with redundancy version (RV), code block concatenation, and interleaving.  The generated codeword is then scrambled, modulated using either QPSK or 16QAM. This is followed by \ac{dft} by means of transform precoding in order to generate the data symbols. Similar to the control channel, \ac{dmrs} symbols are added and transmitted alongside the data symbols in a \ac{pssch} subframe.

All the symbols are then mapped to the sidelink resource grid followed by \ac{scfdma} modulation to create the time domain waveform. The generated time domain waveform is then filtered through a channel and \ac{awgn} noise is added to it.

\subsubsection{Receiver Operations}
For each resource pool as configured in the resource pool selection, the receiver tries to perform a blind decoding of the control information by iterating over all possible cyclic shift values. For each selected cyclic shift, the receiver first corrects the frequency offset, demodulates the \ac{scfdma} time domain symbols to recover the resource grid. This is followed by channel estimation using a cubic interpolation over a pre-specified time and frequency window. The effect of the channel is equalized by dividing the received grid with that of the estimated channel grid. After this, the control symbols are extracted and are then decoded (by performing the inverse operations) to recover the \ac{sci} message. If the \ac{sci} decoding is successful, then the receiver converts the 16 bit CRC checksum into a decimal NXID is used to proceed with decoding the data message. If the decoding is not successful, it means that the shared data is also discarded.

After decoding the \ac{sci} message and recovering the NXID, the receiver proceeds with decoding the data. Similar operations (channel estimation, equalization and turbo decoding) are performed to recover the data block.

\begin{figure*}[]
	\centering
	\includegraphics[width=\textwidth]{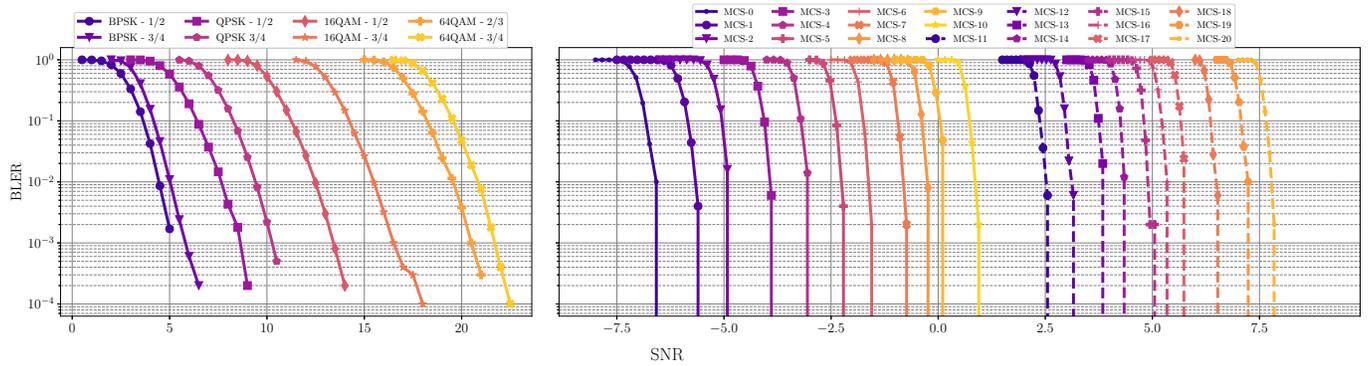}
	\caption{Performance over \ac{awgn} Channel}
	\label{awgn_perf}
\end{figure*} 

\begin{figure*}[htp]
	\centering
	\includegraphics[width=\textwidth]{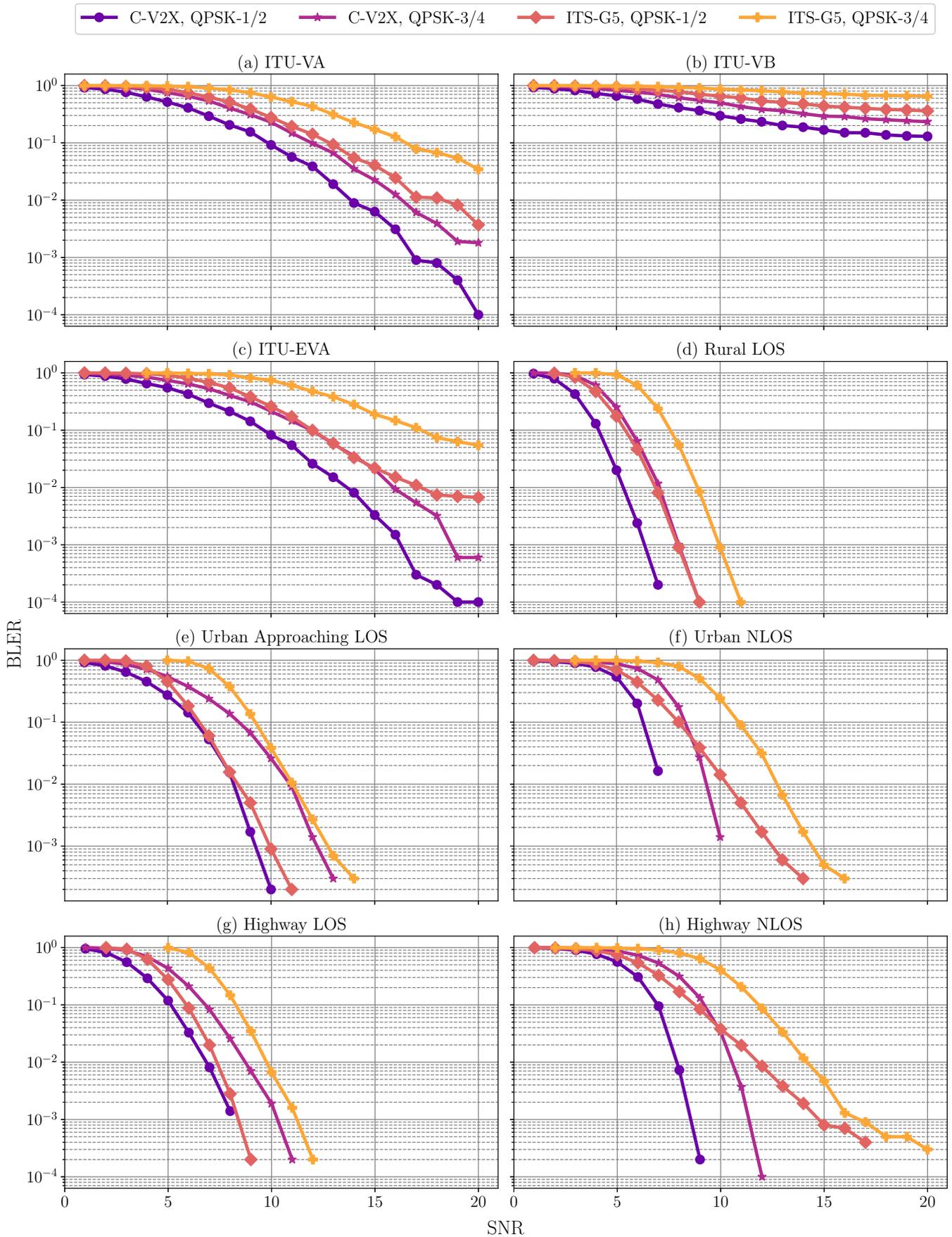}
	\caption{Comparison of C-V2X and ITS-G5 over Fading Channels}
	\label{fading_comp}	
\end{figure*}

\section{Performance Comparison over Fading Channels}
In order to have a fair comparison between \ac{ieee} 802.11p and \ac{cv2x}, some baseline assumptions are required to be made. A packet size of 300 bytes (2400 bits) is assumed which seems to be an acceptable value for safety messages \cite{Khairnar2013, c2c_report2018, Ma2009}. For fairness, only one transmission was assumed even though \ac{cv2x} supports one blind retansmission by default. Since, we limit our analysis to safety messages, we only consider lower order modulation schemes, i.e., QPSK with coding rates 1/2 and 3/4. For \ac{ieee} 802.11p, this corresponds to \ac{mcs} 2 and 3 respectively. In case of \ac{cv2x} which is based on LTE, it is not possible to set the packet size to exactly 300 bytes since the \ac{mcs} has to be selected from a list of predefined values based on the available \acp{prb}. Moreover, the coding rate is always a bit on the lower/higher side due to rate matching and the tail bits added to the turbo encoder. Hence, the configurations in \ref{tab:cv2x_used_mcs} are used for \ac{cv2x} in order to keep the coding rates as close to that of ITS-G5 and not violating the intended packet size too much.

\begin{table}[]
	\centering
	\caption{C-V2X MCS}
	\label{tab:cv2x_used_mcs}
	\resizebox{0.40\textwidth}{!}{%
		\begin{tabular}{@{}ccccc@{}}
			\toprule
			MCS Scheme & MCS Index & TBS & PRBs & \begin{tabular}[c]{@{}c@{}}Effective \\ Coding Rate\end{tabular} \\ \midrule
			QPSK 1/2 & 7 & 2472 & 20 & 0.515 \\
			QPSK 3/4 & 10 & 2664 & 15 & 0.74 \\ \bottomrule
		\end{tabular}%
	}
\end{table}

\subsection{ITU Channel Models}
ITU \cite{IEEE} specifies three different test environments: Indoor office, outdoor-to-indoor pedestrian and vehicular-high antenna. For the vehicular test environment, a low (A) and medium (B) delay spreads have been defined with 6 channel taps and an RMS delay spread of 370 ns and 4000 ns respectively. The Channel A was extended with additional taps to support higher bandwidths. The path delays and gains of these models are outlined in Table \ref{tab:itu_models}


\begin{table}[]
	\centering
	\caption{ITU Channel Models}
	\label{tab:itu_models}
	\resizebox{0.49\textwidth}{!}{%
		\begin{tabular}{@{}lll@{}}
			\toprule
			\multicolumn{1}{c}{\textbf{Model}} & \multicolumn{1}{c}{\textbf{Path Delays (ns)}} & \multicolumn{1}{c}{\textbf{Path Gains (dB)}} \\ \midrule
			ITU - VA & [0, 310, 710,  1090, 1730, 2510] & [0, -1, -0, -10, -15, -20] \\
			ITU - VB & [0, 300, 8900, 12900, 17100, 20000] & [-2.5, 0, -12.8, -10, -25.2, -16] \\
			ITU-EVA & [0, 30, 150, 310, 370, 710, 1090, 1730, 2510] & [0, -1.5, -1.4, -3.6, -0.6, -9.1, -7, -12, -16.9] \\ \bottomrule
		\end{tabular}%
	}
\end{table}


\subsection{\ac{ieee} Tiger Team Channel Models}

During 2007-2010, a total of 35 field trial campaigns were conducted on public roads in US, Germany, Austria, Italy and Australia totalling over 1100 kilometres \cite{Alexander2011a}. These campaigns demonstrated different \ac{v2i} and \ac{v2v} scenarios such as \ac{ima}, \ac{dnpw}, \ac{eebl} and driving across an \ac{rsu}. For each test location, multiple repetitions of a scenario were run transmitting messages at an aggregate of 400 packets/s. For the purpose of measurements, vehicles mounted with Cohda wireless MKI \ac{ieee} 802.11p \ac{dsrc} units with single antenna were used. The channel sounding data captured during the field trials were analysed to obtain delay and Doppler spread characteristics. Using these statistics, a total of 5 channel models were proposed for different scenarios and are outlined in Table \ref{tab:channel_models}.

\begin{table}[]
	\centering
	\caption{V2V Channel Models}
	\label{tab:channel_models}
	\resizebox{0.49\textwidth}{!}{%
		\begin{tabular}{@{}llll@{}}
			\toprule
			\textbf{Scenario} & \textbf{Path Delays (ns)} & \textbf{Path Gains (dB)} & \textbf{Doppler Shift (Hz)} \\ \midrule
			Rural LOS & [0, 83, 183] & [0, -14, -17] & [0, 492, -295] \\
			\begin{tabular}[c]{@{}l@{}}Urban Approaching\\ LOS\end{tabular} & [0,  117, 183, 333] & [0, -8, -10, -15] & [0, 236, -157, 492] \\
			Urban NLOS & [0, 267, 400, 533] & [0, -3, -5, -10] & [0, 295, -98, 591] \\
			Highway LOS & [0, 100, 167, 500] & [0, -10, -15, -20] & [0, 689, -492, 886] \\
			Highway NLOS & [0, 200, 433, 700] & [0, -2, -5 -7] & [0, 689, -492, 886] \\ \bottomrule
		\end{tabular}%
	}
\end{table}

Figure \ref{awgn_perf} shows the \ac{bler} performance over \ac{awgn} channels for both \ac{ieee} 802.11p and \ac{cv2x}. It can be seen that for the considered \ac{mcs} schemes, i.e., QPSK 1/2 and 3/4, \ac{cv2x} provides a performance gain of almost close to 10 dB. This is because of the use of turbo encoder compared to a convolutional encoder that is used in \ac{ieee} 802.11p. Secondly, due to the presence of a higher number of \ac{dmrs} symbols in \ac{cv2x} when compared to \ac{ieee} 802.11p, the noise is also estimated better resulting in more robust channel equalization.

%

Figure \ref{fading_comp} shows the performance comparison over fading channels as outlined in Tables \ref{tab:itu_models} and \ref{tab:channel_models}. For the case of \ac{itu} channel models (a-c), it can be clearly seen that \ac{cv2x} exhibits a gain of almost 4-6 dB. The gain is more pronounced for coding rate 3/4 than 1/2. The performance is almost similar for both \ac{itu}-\ac{va} and \ac{itu}-\ac{eva} channels. This is expected since \ac{itu}-\ac{eva} is just an extension of \ac{itu}-\ac{va} channel model with more paths. However, both the technologies perform very poorly for \ac{itu}-\ac{vb}. This is due to the very large delay spread (20000 ns) that is way greater than the \ac{cp} length and thereby causing high inter-symbol interference. However, it can be noted that \ac{cv2x} still performs a bit better than \ac{ieee} 802.11p.

Figures \ref{fading_comp} (d-h) show the performance comparison for Tiger team channel models. It can be seen that \ac{cv2x}, in general fares better than \ac{ieee} 802.11p for all scenarios with gains ranging from 0-5 dB with the exception of model (e) where the performance of both the technologies is almost identical. The performance of \ac{ieee} 802.11p QPSK 1/2 is similar to that of \ac{cv2x} QPSK 3/4 for model (d). It can also be seen that \ac{cv2x} performs better for \ac{nlos} scenarios, especially for scenario (h) where the vehicles speeds are higher. This shows that \ac{cv2x} is better equipped to handle high speed scenarios which in turn is due to the higher number of \ac{dmrs} symbols thereby resulting in better channel estimation performance.

\section{Conclusions \& Future Work}
In this work, we evaluated the link level performance of the two candidate technologies for \ac{v2x} communication, namely \ac{ieee} 802.11p and \ac{cv2x} for different channel models. The considered channel models include those from the ITU (VA, VB and EVA) and the \ac{dsrc} channel models from the \ac{ieee} Tiger team that were developed after extensive field trials. Two \ac{mcs} schemes - QPSK 1/2 and QPSK 3/4 were considered for the evaluation for a packet size of 300 bytes. The results show that \ac{cv2x} outperforms \ac{ieee} 802.11p for almost all of the considered channel models with a gain ranging from 0-5 dB. Moreover, it is also clear from the results that \ac{cv2x} performs better at higher vehicle speeds. This better performance of \ac{cv2x} can be attributed to the use of turbo encoder and the better channel estimation mechanism that makes use of a higher number of \ac{dmrs} symbols.

\section*{Acknowledgements}
Part of this work has been performed in the framework of the BMVI project Connected Vehicle (V2X) of Tomorrow (ConVeX). The authors would like to acknowledge the contributions of their colleagues, although the views expressed are those of the authors and do not necessarily represent the project.

\bibliographystyle{IEEEtran}
\bibliography{link_level_vtc}

\end{document}